\documentclass[%
 aip,
 amsmath,amssymb,
 reprint,%
nofootinbib
]{revtex4-1}

\usepackage[pdftex]{graphicx}
\usepackage{dcolumn}
\usepackage{bm}

\usepackage[utf8]{inputenc}
\usepackage[T1]{fontenc}
\usepackage{mathptmx}
\usepackage{etoolbox}
\usepackage{color}

\makeatletter
\def\@email#1#2{%
 \endgroup
 \patchcmd{\titleblock@produce}
  {\frontmatter@RRAPformat}
  {\frontmatter@RRAPformat{\produce@RRAP{*#1\href{mailto:#2}{#2}}}\frontmatter@RRAPformat}
  {}{}
}%
\makeatother
\begin{document}

\preprint{AIP/123-QED}

\title{Exploration of stable atomic configurations in graphene-like BCN systems by Bayesian optimization}

\author{T. Hara}
\affiliation{%
 Interdisciplinary Graduate School of Engineering Sciences, Kyushu University, Fukuoka 816-8580, Japan
}%

\author{A. Kusaba}
 \email{kusaba@riam.kyushu-u.ac.jp}
\author{Y. Kangawa}
\affiliation{%
 Research Institute for Applied Mechanics, Kyushu University, Fukuoka 816-8580, Japan
}%

\author{T. Kuboyama}
\affiliation{
Computer Centre, Gakushuin University, Toshima-ku, Tokyo 171-8588, Japan
}

\author{D. Bowler}
\affiliation{
 London Centre for Nanotechnology, UCL, 17-19 Gordon St, London, WC1H 0AH, UK
}

\author{K. Kawka}
\author{P. Kempisty}%
\affiliation{%
Institute of High Pressure Physics, Polish Academy of Sciences, Sokolowska 29/37, 01-142 Warsaw, Poland
}%

\date{\today}

\begin{abstract}
h-BCN is an intriguing material system where the bandgap varies considerably depending on the atomic configuration, even at a fixed composition. 
Exploring stable atomic configurations in this system is crucial for discussing the energetic formability and controllability of desirable configurations. 
In this study, this challenge is tackled by combining first-principles calculations with Bayesian optimization. 
An encoding method that represents the configurations as vectors, while incorporating information about the local atomic environments and domain knowledge, is proposed for the search. 
The proposed encoding method proved effective in the search, resulting in the discovery of two interesting and stable semiconductor configurations. 
Furthermore, the optimization behavior is discussed through principal component analysis, confirming that the ordered BN network and the C configuration features are well embedded in the search space. 
While our approach provided a tailored encoding for the h-BCN system in this study, it holds promise for broader application to other materials by adapting the domain knowledge matrix to each target system.
\end{abstract}

\maketitle

\section{Introduction}
The h-BCN system~\cite{rubio2010nanoscale,song2012binary,gong2014direct,miyamoto1994chiral,weng1995synthesis}, consisting of alloys that combine conductive graphene~\cite{wei2009synthesis,schwierz2010graphene,bai2010graphene,zhou2010electronic,brito2012b}, which lacks a bandgap, and insulating \mbox{h-BN}~\cite{rubio1994theory,nag2010graphene,xu2013graphene}, characterized by its wide bandgap, is gaining attention as a promising material for post-silicon electronic devices. 
Notably, it has been reported that the atomic configuration in h-BCN alloys can significantly influence its electronic properties, leading to variations in the bandgap that can range from metallic to semiconducting behavior even with a fixed composition~\cite{liu1989atomic,azevedo2006structural,zhu2011interpolation}.
Understanding the relationship between atomic configurations and energetic stability is crucial for elucidating and controlling the growth processes to achieve designed configurations. 
Thus, a key question is: what configurations are more likely to form energetically?
Additionally, investigating mixing energy is also important to assess the potential for segregation or domain formation\cite{ci2010atomic,peng2012tunable}, phenomena that might cause spatial modulation of material properties.

In recent years, a synthetic method using bis-BN cyclohexane (B$_2$N$_2$C$_2$H$_{12}$) as a precursor molecule has been developed, enabling the formation of monolayers based on B-C-N-B-C-N rings as a structural unit~\cite{beniwal2017graphene}.
Such a synthetic approach, utilizing molecular motifs, is expected to enhance the controllability of the configuration by imposing restrictions on it.
Additionally, research has been conducted on a synthetic method involving the reaction of boron trichloride (BCl$_3$) with polyacrylonitrile ([CH$_2$CHCN]$_n$) to crosslink the polyacrylonitrile chain with boron~\cite{kawaguchi1993synthesis}, as well as a method using acrylonitrile monomers (CH$_2$CHCN)~\cite{kawaguchi1996syntheses,kawaguchi1999synthesis}. 
The latter method is considered to allow access to configurations with greater flexibility.
Furthermore, a technique has been reported for converting selected areas of an h-BN sheet into graphene~\cite{kim2015catalytic,kim2020effect}, with one example involving the formation of graphene quantum dots~\cite{kim2019planar}.

While first-principles calculations can be used to obtain the mixing energy between graphene and h-BN, the vast number of possible configurations makes exhaustive computation impractical. 
For example, for the composition and periodicity settings demonstrated in this study, the number of potential configurations reaches the order of $10^6$. 
To overcome this combinatorial explosion, techniques for efficient structure sampling using quantum annealing~\cite{nawa2023quantum,lin2025determination}, Bayesian optimization~\cite{seko2020prediction,kusaba2022exploration,ono2022optimization,iwasaki2022efficient}, replica exchange Monte Carlo method~\cite{kasamatsu2023configuration} and Metropolis–Hastings algorithm~\cite{du2023machine}, the latter two combined with neural network potentials and force fields~\cite{behler2007generalized,schutt2021equivariant}, have been proposed. 
These techniques have been applied to systems such as barrier materials in magnetic tunnel junctions~\cite{nawa2023quantum}, scandium-doped barium zirconate~\cite{lin2025determination}, oxygen-deficient perovskites~\cite{seko2020prediction}, semiconductor surface reconstructions~\cite{kusaba2022exploration,du2023machine,kawka2024augmentation}, two-dimensional copper-based systems~\cite{ono2022optimization}, Heusler alloys~\cite{iwasaki2022efficient}, materials with high Curie temperatures~\cite{iwasaki2024autonomous}, half-metallic materials with a B2 structure~\cite{iwasaki2024autonomousb2}, and metallic oxide alloys~\cite{kasamatsu2023configuration}.

In this study, we explore stable configurations of h-BCN using a combination of Bayesian optimization and first-principles calculations. 
Bayesian optimization is capable of efficiently minimizing an objective function with a relatively small number of trials. Therefore, among the various available techniques, it is considered particularly well suited to our task, where the objective is energy minimization and the evaluation requires time-consuming first-principles calculations.
In this process, the configuration search via Bayesian optimization is made feasible for the two-dimensional semiconductor material system h-BCN by utilizing a system-tailored encoding method that does not increase dimensionality. 
Furthermore, we discuss the optimization behavior through visualization based on dimensionality reduction of the search space.

\section{Methods}
\subsection{Search Space}
The h-BCN configuration search in this study focuses on the range defined by a (3$\times$3) periodicity and a composition ratio of h-BN/graphene = 2, as shown in Fig.~1(a). 
This provides a reasonable number of candidate structures for the demonstration of our proposed approach. 
Under this setting, there are six B atoms, six N atoms, and six C atoms within the (3$\times$3) cell, which can be freely arranged across 18 sites. 
Fixing one C atom at the \textit{site-1} does not compromise generality. 
In this case, the total number of candidate structures within the search space is given by ${}_{17}C_{6} \times {}_{11}C_{6} \times {}_{5}C_{5}$ = 5,717,712.

\begin{figure}
\includegraphics[width=7.1cm]{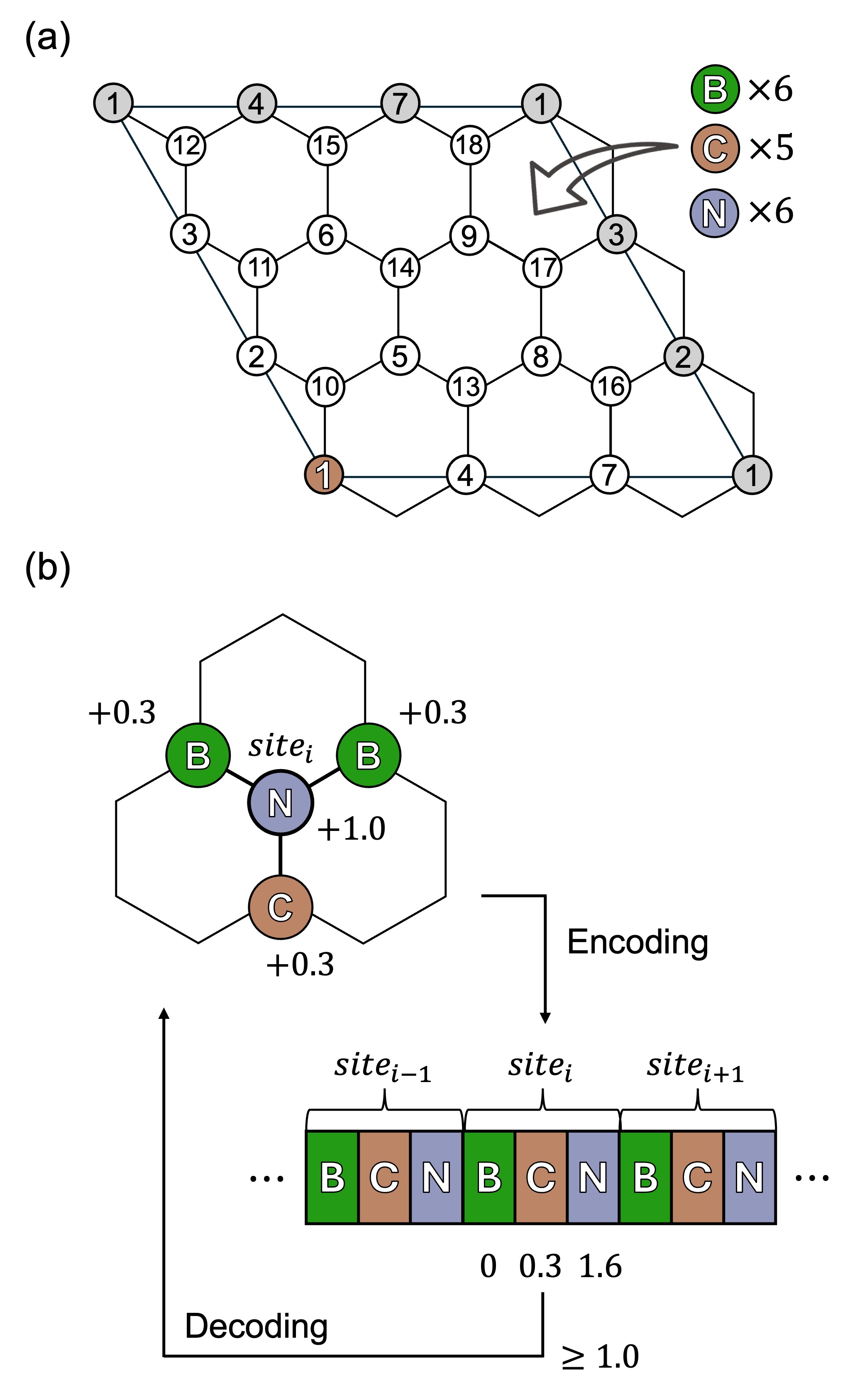}
\caption{(a) Search space: All possible configurations obtained by placing 6 B, 5 C, and 6 N atoms at the \textit{site-2} to the \textit{site-18}. (b)~Domain Knowledge-assisted Neighbor-Atom (DKNA) encoding: The atomic occupancy at the \textit{site-i} as well as information about the nearest neighboring atoms are represented by a 3-element vector. These 3-element vectors are then concatenated to represent the hexagonal network.}
\end{figure}

\subsection{Encoding}
In standard Bayesian optimization based on a Gaussian kernel, a configuration (i.e., a hexagonal network) needs to be represented as a vector. 
In one-hot encoding~\cite{rodriguez2018beyond}, the \textit{site-i} has three values ($x_{i\text{-}B}$, $x_{i\text{-}C}$, $x_{i\text{-}N}$) corresponding to the three elements. 
When element Z (= B, C, or N) is present at the \textit{site-i}, the value $x_{i\text{-}Z}$ corresponding to Z is set to 1, while the other values are set to 0, representing the atomic occupancy at the \textit{site-i}. 
In this case, a configuration is represented by a vector of 54 dimensions, corresponding to 18~sites $\times$ 3~elements. 
However, such one-hot encoding does not incorporate information about neighboring sites in the hexagonal network. 
To address this, the one-hot vector is modified as follows.
\begin{equation}
\begin{pmatrix}
x_{i\text{-}B} \\
x_{i\text{-}C} \\
x_{i\text{-}N}
\end{pmatrix}
+
w
\begin{pmatrix}
0 & 0 & 1 \\
0 & 1 & 0 \\
1 & 0 & 0
\end{pmatrix}
\begin{pmatrix}
n_{i\text{-}B} \\
n_{i\text{-}C} \\
n_{i\text{-}N}
\end{pmatrix}
\end{equation}
Here, the second term, namely the modification term, consists of the weight $w$ (set to 0.3 in this study), a domain knowledge matrix, and the neighboring atom counts $n_{i\text{-}Z}$ (i.e., the number of element Z present among the three nearest neighboring sites of \textit{site-i}). 
In this case, in order to incorporate domain knowledge that B-N bonding is expected to contribute to stabilization, a matrix was designed to swap the neighboring atom counts between B and N. 
This matrix is intended to assist the co-occurrence of neighboring B and N atoms by utilizing expressions such as $x_{i\text{-}B} + w n_{i\text{-}N}$ (or $x_{i\text{-}N} + w n_{i\text{-}B}$) for the optimization. In other words, this expression increases when a B (or N) atom is placed at \textit{site-i} and N (or B) atoms are located around it (see Fig.~1(b)).
Hereafter, this is referred to as \textit{Domain Knowledge-assisted Neighbor-Atom (DKNA) encoding}.
In the case of $3w<1$, decoding, or converting the vector back to a configuration, can be uniquely performed by assigning the element with the highest value, which is always 1 or higher, to the corresponding site. 
This is because, under this condition, the maximum possible contribution from the three neighboring sites is less than the one-hot encoding value of 1.0 for the central site.

\subsection{Sampling}
The sampling of structures (i.e., atomic configurations) was carried out based on Bayesian optimization~\cite{snoek2012practical}, which was set up to minimize the total energy of the system. 
First, two randomly sampled structures were used as the initial data. 
Their total energies were calculated based on density functional theory (DFT). 
Using this training data, a Bayesian linear model was trained. 
This model predicts the relaxed DFT total energy from input vectors encoded using either one-hot or DKNA encoding. 
Based on the predicted values and their uncertainties, one structure to be examined next is proposed using Thompson sampling~\cite{chapelle2011empirical}. 
This new sample is then evaluated by DFT, added to the training data, and the model is updated. 
This loop is repeated to efficiently sample stable structures. 
The dimension of the random feature map~\cite{rahimi2007random}, one of the hyperparameters of the model, was set to 3000, and other hyperparameters were updated in each trial based on the maximization of type-II likelihood~\cite{rasmussen2003gaussian}. 
In this study, we employed the implementation provided by the Python library PHYSBO~\cite{motoyama2022bayesian}, and for further details on the principles, please refer to the literature~\cite{ueno2016combo}.

\subsection{DFT Calculations}
For the stability evaluation of the samples, DFT calculations were performed using the Quantum ESPRESSO program package\cite{giannozzi2009quantum,giannozzi2017advanced}, which is based on plane-wave basis sets and pseudopotentials, with the following settings. 
The exchange and correlation energies were treated using the generalized gradient approximation (GGA) with the Perdew–Burke–Ernzerhof (PBE) exchange-correlation functional~\cite{perdew1996generalized}. 
Projector-augmented wave (PAW) pseudopotentials were used~\cite{kresse1999ultrasoft}. 
The cutoff energies for the wave functions and the charge densities were set to 30 and 120 Ry, respectively. 
The k-point sampling was carried out with a 3$\times$3$\times$1 grid (or 3$\times$5$\times$1 for the armchair type, as mentioned later). 
Structural optimization was performed with periodic boundary conditions and variable lattice constants, using convergence criteria of $10^{-5}$~Ry for energy and $10^{-4}$~Ry/Bohr for forces.

\section{Results}
Figure~2 shows the two initial data points and the subsequent sampling history of 1,000 structures using Bayesian optimization for each of the two encoding cases. 
Since structures where the DFT calculation did not converge were skipped, the total number of displayed data points is less than 1,002. 
The DFT total energy of the samples is expressed as the following mixing energy:
\begin{equation}
  E_{mix} = E_{h\text{-}BCN} - \frac{1}{3}E_{graphene} - \frac{2}{3}E_{h\text{-}BN},
\end{equation}
where $E_{h\text{-}BCN}$, $E_{graphene}$, $E_{h\text{-}BN}$ are the DFT total energies of h-BCN, graphene, and h-BN, respectively, calculated with the same cell size, with the latter two being constants. 
Please recall that our objective was to search for configurations with lower mixing energy. 
Similar oscillatory behavior along trials is observed for both one-hot encoding and DKNA encoding. 
However, the trends differ between the two cases: while the mean value for one-hot encoding remains almost flat, a decreasing trend in mixing energy is observed as the sampling progresses in the case of DKNA encoding.

\begin{figure}
\includegraphics[width=8.5cm]{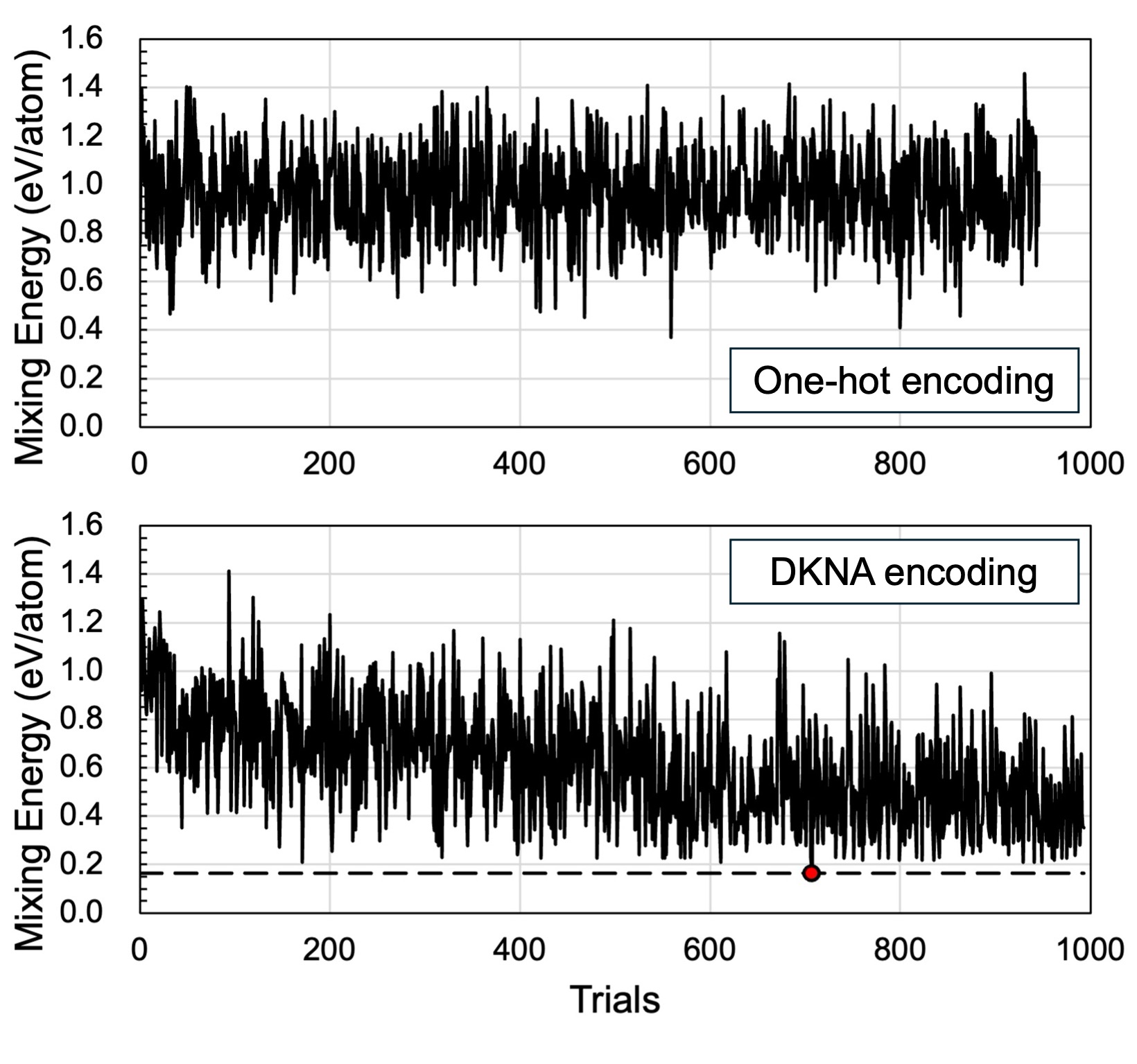}
\caption{History of structure sampling by Bayesian optimization. The red marker indicates the discovered zigzag-type stripe configuration.}
\end{figure}

The most stable structure among the sampled ones with the DKNA encoding (red marker in Fig.~2) has a mixing energy of 0.164~eV/atom.
This structure is shown in Fig.~3(a). 
It is a configuration where carbon atoms are introduced in a striped pattern on an ordered h-BN configuration, which we will refer to as the zigzag type. 
Although we explored within the periodicity of (3$\times$3) this time, a different stripe configuration within the same composition can be considered if a different periodicity is assumed. 
The configuration shown in Fig.~3(b) also has carbon stripes introduced on an ordered h-BN, but in a different direction from the zigzag type, and we will refer to this as the armchair type. 
The mixing energy of the armchair type is 0.158~eV/atom, and both of these stripe configurations have very similar stability. 
The relationship between other sampled configurations and optimization behavior will be discussed in detail later.

\begin{figure}
\includegraphics[width=8.5cm]{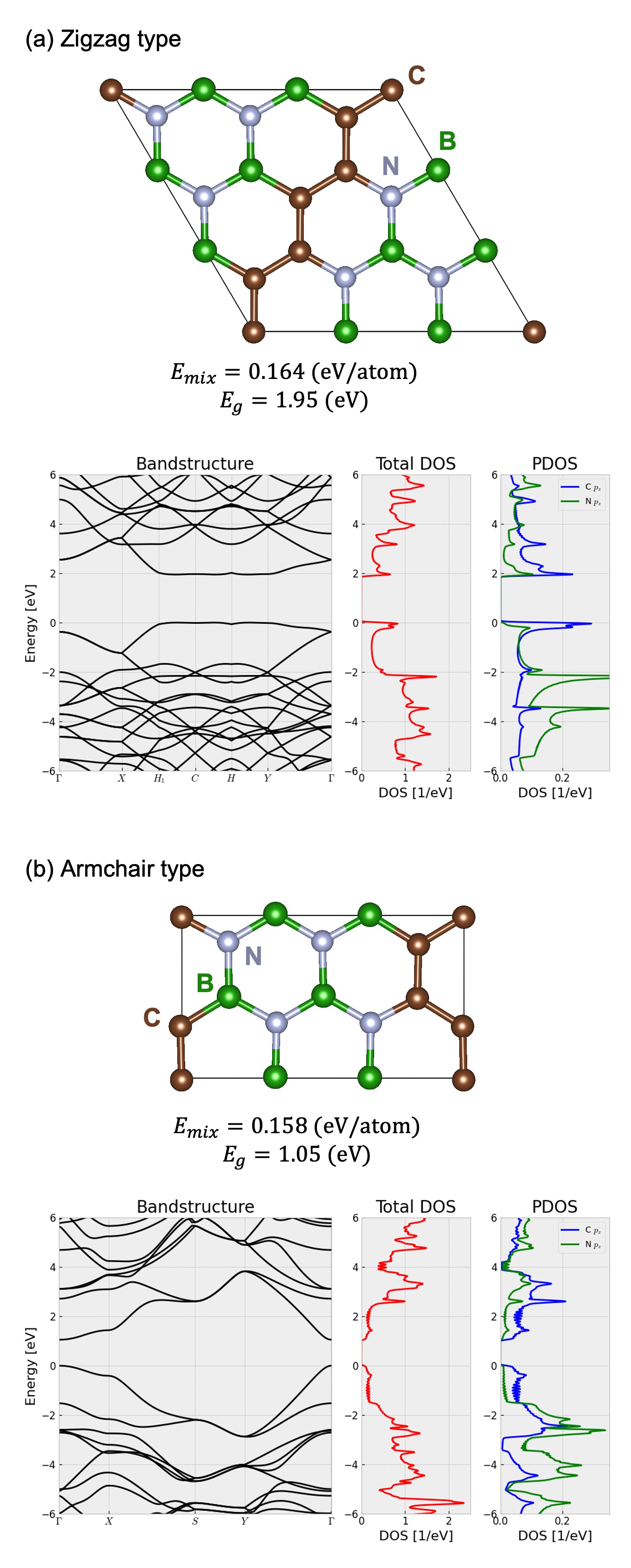}
\caption{Configurations and band structures of (a) the zigzag type (identified as the most stable among the samples) and (b) the armchair type. The mixing energies and bandgaps are also indicated.}
\end{figure}

To further investigate the two discovered stripe structures, detailed analyses were conducted.
The resulting band structure (dispersion relation), total density of states (DOS), and the projected DOS (PDOS) for the $p_z$ carbon and nitrogen orbitals are shown in Fig.~3.
The most notable differences in the valence band between the zigzag and armchair structures appear in the PDOS for the $p_z$ carbon and $p_z$ nitrogen orbitals, which play a substantial role in determining the bandgap width. 
Calculations indicate a bandgap of 1.95~eV for the zigzag structure and 1.05~eV for the armchair structure.
These values are consistent with the range of bandgap values, which depend on atomic configuration, reported in previous studies for the same composition: 0.98–1.82~eV at PBE level (corresponding to 1.33–2.68~eV at HSE level)~\cite{zhu2011interpolation}, 1.50~eV at HSE level~\cite{beniwal2017graphene}, and 1.8~eV at PBE level~\cite{raidongia2010bcn}.

\section{Discussion}
\subsection{Discovered Stripe Configurations}
The stripe-type semiconductor configurations discovered in this study are energetically more stable and exhibit more ordered patterns compared to models with the same composition and periodicity proposed in two previous studies. 
In these studies, (3$\times$3) models with mixing energies of 0.44~eV/atom~\cite{beniwal2017graphene} and 0.28~eV/atom~\cite{raidongia2010bcn} were reported, respectively.
This signifies the achievement of the demonstration's objective. 
Note that in previous studies, configuration searches were not conducted in a search space containing a vast number of candidates as in this study; only a limited number of models were investigated.
Beyond our (3$\times$3) periodicity setting, however, our models are not the most stable among the same composition. 
While less stable models with mixing energies of 0.68~eV/atom~\cite{azevedo2006structural}, 0.55~eV/atom~\cite{azevedo2006structural}, 0.31~eV/atom~\cite{beniwal2017graphene}, and 0.21~eV/atom~\cite{beniwal2017graphene} were reported in previous studies, more stable models with mixing energies of 0.12~eV/atom~\cite{zhu2011interpolation} and 0.11~eV/atom~\cite{zhu2011interpolation} were also reported.
These more stable models from the previous study~\cite{zhu2011interpolation} have configurations related to our stripe configurations, referred to as carbon paths.
But, the C and BN regions in these configurations are more separated, or less mixed, than in our stripe configurations. 
Furthermore, the stripe configurations are consistent with the stabilization rule discussed in a previous study~\cite{beniwal2017graphene}, which suggests that stable \mbox{h-BCN} structures tend to maximize the number of B–N and C–C bonds.
Within the search space of (3$\times$3) periodicity, a structure where one carbon ring was introduced into an ordered h-BN configuration is included. 
Although this structure had the same number of B–N and C–C bonds as the zigzag-type structure, it was confirmed to be slightly less stable (0.173~eV/atom). 
Also in the different composition BC$_2$N, a zigzag-type model has been similarly considered, which can be represented within the (2$\times$2) periodicity~\cite{liu1989atomic}.
Thus, the stripe structures discovered through Bayesian optimization are considered plausible and energetically possible to form, as they do not contradict the qualitative trends and discussions from prior studies.

On the other hand, these stripe structures are not configurations accessible through the synthesis method based on molecular motifs that has proven effective in forming h-BCN monolayers~\cite{beniwal2017graphene}. 
In contrast, in a synthesis method involving polymer cross-linking, a structure similar to the stripe configurations has been suggested, although its composition is different~\cite{kawaguchi1996syntheses}. 
Additionally, the result that the two stripe structures have nearly equal stability despite an approximately 0.9 eV difference in their bandgaps could pose a challenge from a controllability perspective. 
In other words, the two stripe structures could coexist.
Such an issue has been previously recognized in the h-BCN system~\cite{zhu2011interpolation}. 
The two points of whether their coexistence would result in stable configurations and whether interpolated bandgaps can be obtained are uncertain, and further studies using larger computational cells would be necessary.

\subsection{Optimization  Behavior}
\begin{figure*}
\includegraphics[width=\textwidth]{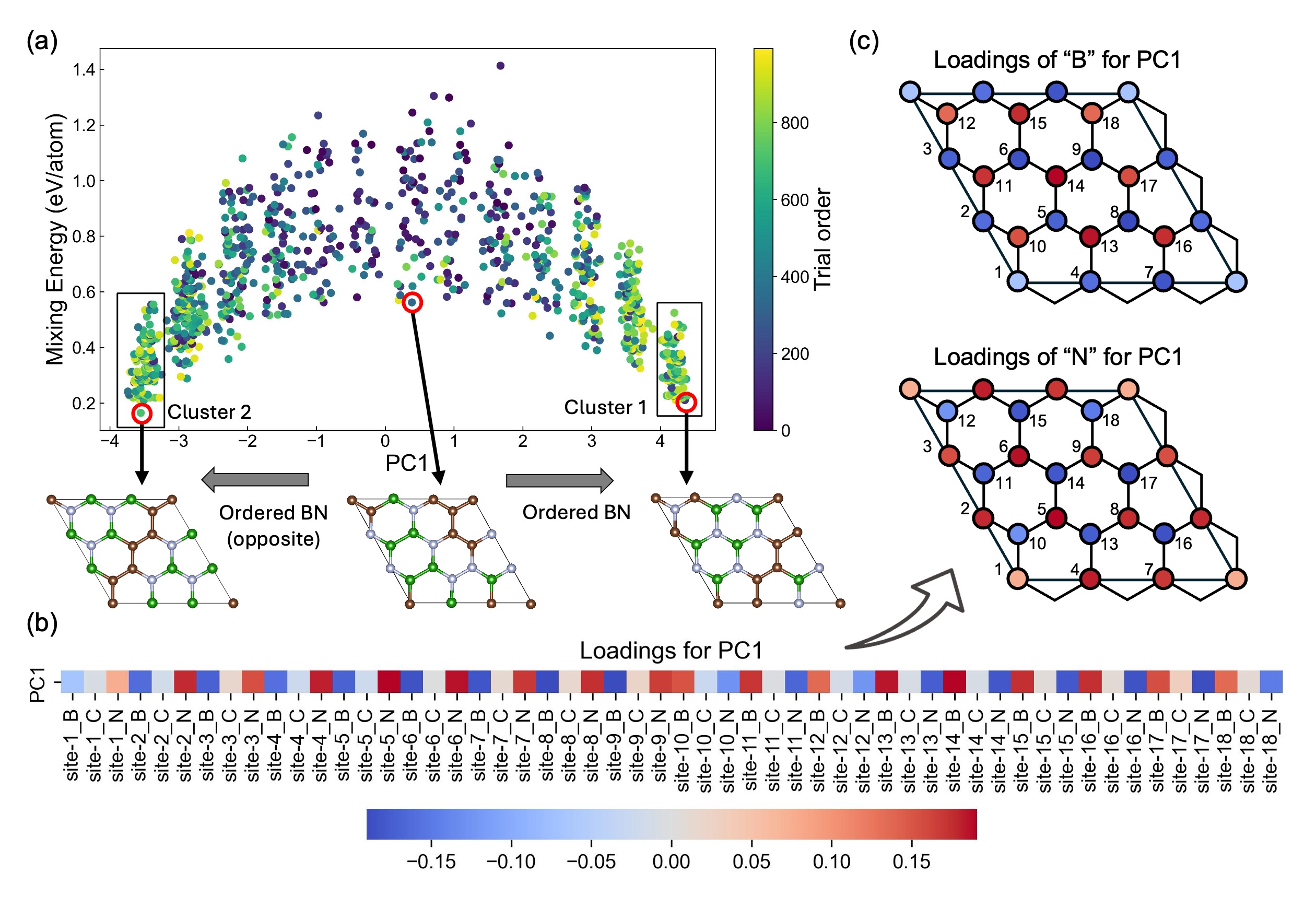}
\caption{Principal component analysis of the structures sampled using the DKNA encoding. (a) Plot of the 1st principal component (PC1) versus the mixing energy, with markers colored by trial order. (b) Heatmap of the loadings for PC1. (c) Loadings for PC1 displayed on the sites for each element, using the same colors as in (b).}
\end{figure*}
Next, we perform principal component analysis (PCA) on the sampled structures to discuss the optimization behavior. 
Recall that the search space is an 18-site by 3-element dimensional space, where structures are embedded based on DKNA encoding. 
PCA reveals the direction in which the samples are most dispersed in this space as the 1st principal component (PC1). 
Figure~4(a) shows a plot of the mixing energy of the samples along the PC1 axis. 
It can be observed that as the absolute value of the PC1 increases, the mixing energy decreases. 
Since PC1 is an axis in the search space, this indicates that the samples are embedded in a gradient from unstable to stable structures in the search space, suggesting that the DKNA encoding functions well as a feature descriptor.
Additionally, the color of the markers corresponds to the sampling order. 
It can be seen that the Bayesian optimization progressed from smaller to larger absolute values of PC1, implying that the Bayesian linear model appropriately captured the relationship between structure and stability.

Additionally, to further investigate the relationship between PC1 and the configurations, we present the loadings of the original variables for PC1 in Fig.~4(b). 
PC1 is a linear combination of the original variables in the search space, and the loadings can be considered as the lengths of the projections of the original basis vectors onto the PC1. 
Figure~4(c) displays these loadings mapped onto the crystal sites for easier interpretation. 
It can be observed that the sites with positive loadings shown in red and those with negative loadings shown in blue are arranged alternately, corresponding to an ordered BN network. 
Specifically, the larger the PC1 value, the more B atoms are located at the red sites for the B loadings, and the more N atoms are located at the red sites for the N loadings. 
Conversely, the smaller the PC1 value, the more the respective elements are located at the blue sites. 
In fact, structures with disordered BN configurations were sampled when the absolute value of PC1 was small, and as the absolute value of PC1 increased, structures with ordered BN configurations were sampled. 
The B and N sites were reversed depending on the sign of PC1.

\begin{figure}
\includegraphics[width=7.5cm]{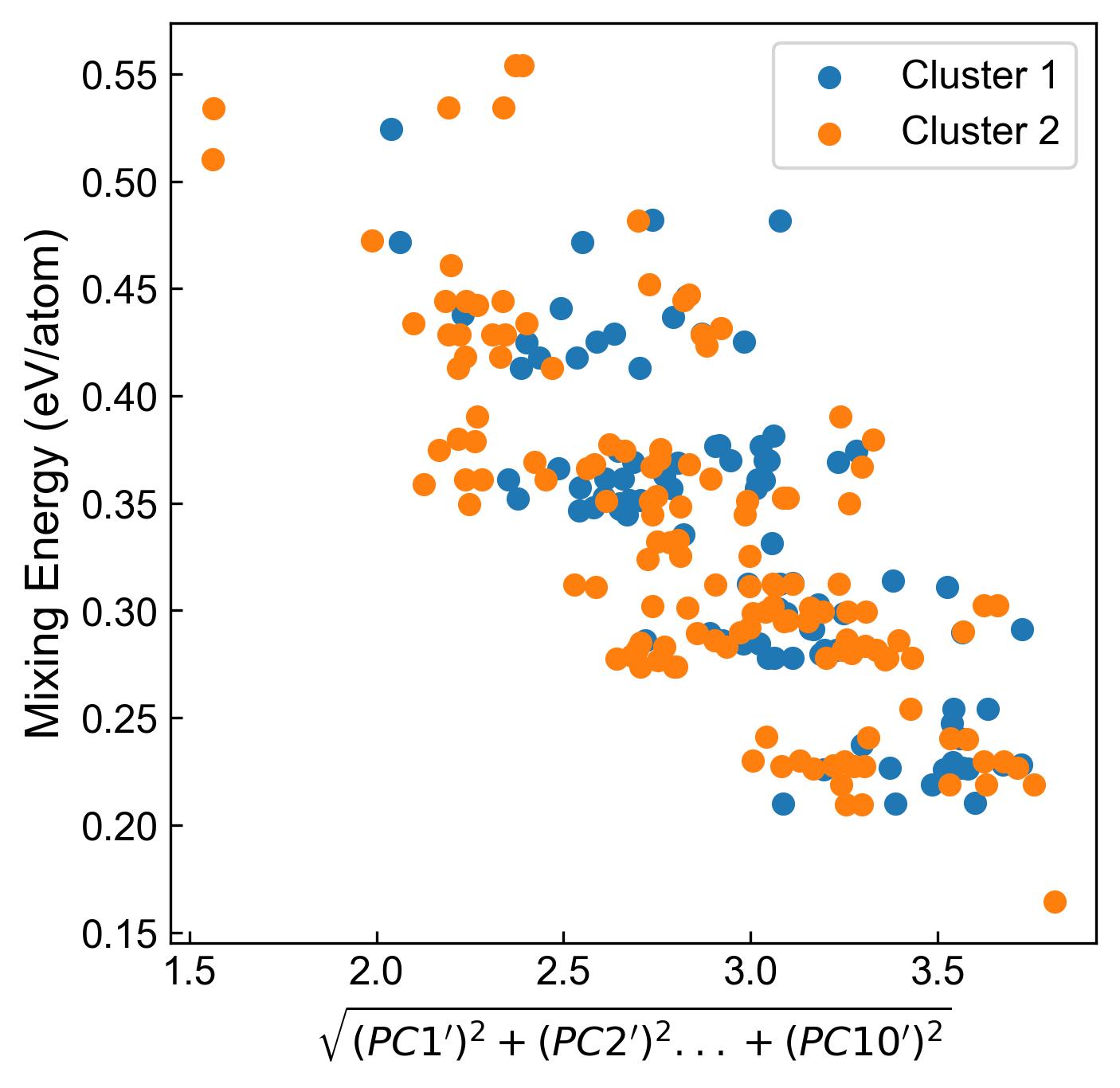}
\caption{Principal component analysis performed separately on the structures within the cluster 1 and the cluster 2 in Fig.~4(a).}
\end{figure}

In the PCA of all samples, the optimization behavior of the C configuration could not be clearly observed. 
This is likely because the ordering of the BN network is more dominant in stabilizing the structures. 
Therefore, an analysis excluding the influence of the BN network's ordering is necessary. 
The structures in clusters 1 and 2 of Fig.~4(a) possess ordered BN networks, and they primarily reflect the differences in stability due to the C configuration. 
We extracted these clusters separately and performed PCA on them individually. 
While the stability of the BN network could be effectively explained in one dimension, the stability of the C configuration is represented in a higher-dimensional space. 
Figure~5 illustrates the case of a 10-dimensional space, where the mixing energy decreases as the Euclidean distance from the center increases. 
This confirms that, for the stability of the C configuration as well, the DKNA encoding functions appropriately.

\section{Conclusions}
In summary, in this study, stable configurations of h-BCN were explored using a combination of Bayesian optimization and first-principles calculations. 
The optimization did not work effectively with one-hot encoding, which led to the need for a more advanced encoding method. 
To address this, a DKNA encoding was proposed, which describes the local environment of crystal sites and incorporates domain knowledge, providing an encoding method without increasing the dimensionality compared to one-hot encoding. 
Using the proposed method, the optimization of the configurations was effectively achieved. 
The optimization behavior was discussed through PCA, and it was confirmed that the features of the BN network and the C configuration were captured. 
One of the h-BCN configurations discovered during the search exhibited qualitatively similar patterns to conventional models and was more stable within the same periodicity. 
Furthermore, they were consistent with a previously discussed stabilization rule. 
Thus, the proposed configuration search method was confirmed to be practically functional.
While this study focused on the exploration based on energetic stability, future work will extend the approach to thermodynamic stability under growth conditions, using methods such as the Ising model and cluster expansion.

\begin{acknowledgments}
This work was partially supported by JSPS KAKENHI (grant numbers JP20K15181, JP23H03461, JP24K17619, JP24H00432); Collaborative Research Program of Research Institute for Applied Mechanics, Kyushu University; and Diversity and Super Global Training Program for Female and Young Faculty (SENTAN-Q), Kyushu University. 
The computation was carried out using the computer resource offered under the category of General Projects by Research Institute for Information Technology, Kyushu University; and under computational allocation No. G96-1904 by Interdisciplinary Centre for Mathematical and Computational Modelling, University of Warsaw.
\end{acknowledgments}

\appendix
\section{Settings for Detailed Analyses}
Detailed analyses of the two discovered stripe structures were conducted under the following settings.
For the structural relaxation and electronic structure calculations, the Vienna Ab initio Simulation Package (VASP), a plane-wave DFT software~\cite{kresse1993ab,kresse1993ab2,kresse1996efficiency,kresse1996efficient}, was employed. 
The calculations were performed using the GGA with the PBE exchange-correlation functional. 
An energy cutoff of 330~eV was used to ensure the accuracy of the plane-wave basis set. 
The Brillouin zone was sampled using 9$\times$9$\times$1 and 9$\times$15$\times$1 k-point meshes for relaxation and denser 49$\times$49$\times$1 and 49$\times$63$\times$1 meshes for density of states (DOS) calculations of the zigzag and armchair structures, respectively.
Convergence criteria for the self-consistent field calculations were set to an energy difference of $10^{-8}$~eV, with a maximum force tolerance of $10^{-4}$~eV/\AA~ for the atomic positions.

%

\end{document}